\documentclass[prl,twocolumn,showpacs,superscriptaddress,preprintnumbers,amssymb]{revtex4}
\usepackage{graphicx}
\usepackage{dcolumn}
\usepackage{bm}
\usepackage{wasysym}

\newcommand{\beq}{\begin{equation}}
\newcommand{\eeq}{\end{equation}}

\begin{document}
\title{Spin Liquid Phases for Spin-1 systems on the Triangular lattice}


\author{Cenke Xu}
\affiliation{Department of Physics, University of California,
Santa Barbara, CA 93106}

\author{Fa Wang}
\affiliation{Department of Physics, Massachusetts Institute of
Technology, Cambridge, Massachusetts 02139}

\author{Yang Qi}
\affiliation{Instutute for Advanced Study, Tsinghua University,
Beijing 100084, China}

\author{Leon Balents}
\affiliation{Department of Physics, University of California,
Santa Barbara, CA 93106} \affiliation{Kavli Institute for
Theoretical Physics, University of California, Santa Barbara, CA,
93106}

\author{Matthew P. A. Fisher}
\affiliation{Department of Physics, University of California,
Santa Barbara, CA 93106}

\begin{abstract}

Motivated by recent experiments on material
$\mathrm{Ba_3NiSb_2O_9}$, we propose two novel spin liquid phases
($A$ and $B$) for spin-1 systems on a triangular lattice. At the
mean field level, both spin liquid phases have gapless fermionic
spinon excitations with quadratic band touching, thus in both
phases the spin susceptibility and $\gamma = C_v/T$ saturate to a
constant at zero temperature, which are consistent with the
experimental results on $\mathrm{Ba_3NiSb_2O_9}$. On the lattice
scale, these spin liquid phases have $\mathrm{Sp(4)\sim SO(5)}$
gauge fluctuation; while in the long wavelength limit this Sp(4)
gauge symmetry is broken down to $ \mathrm{U(1)} \times Z_2$ in
type $A$ spin liquid phase, and broken down to $Z_4$ in type $B$
phase. We also demonstrate that the $A$ phase is the parent state
of the ferro-quadrupole state, nematic state, and the noncollinear
spin density wave state.

\end{abstract}

\date{\today}

\maketitle

A quantum spin liquid (QSL) is a ground state of an insulating
magnet with vanishing static local moments and exotic emergent
excitations.\cite{balents10:_spin} Within spin wave theory for the
simplest Heisenberg Hamiltonians, quantum fluctuations rapidly
decrease with increasing spin quantum number $S$, so it is often
believed that QSLs may occur only in the extreme case of $S=$1/2
spins.  Indeed, the most promising empirical QSL materials are
comprised of spin-1/2
moments\cite{kappa1,kappa2,131dmit1,131dmit2,131dmit3,kagomematerial1}.
However, when the Hamiltonian deviates from the Heisenberg form,
quantum effects can be enhanced also for higher spin, leading to
ground states beyond the usual magnetically ordered ones.
Theoretically, biquadratic and other higher order exchange terms
have been argued to favor multipolar ordered and QSL states in
particular materials, such as the triangular lattice spin-1 magnet
$\mathrm{NiGa_2S_4}$
\cite{nigas2,balentstrebst,senthilni,nitheory1,nitheory2}\ and
certain ordered double perovskites\cite{chen2010exotic}.  Quite
unexpectedly, recent experiments have evidenced QSL behavior in
the spin-1 magnet $\mathrm{Ba_3NiSb_2O_9}$, with spins residing on
triangular lattices with AB stacking.\cite{banisbo} Although the
Curie-Weiss temperature of this material is $\theta_{CW} \sim
-75\mathrm{K}$, no magnetic ordering or phase transition was
observed down to a temperature of $0.35K$, approximately 200 times
lower than $|\theta_{CW}|$.  The low temperature thermodynamics of
this material is strikingly similar to that of the geometrically
similar spin-1/2 organic triangular lattice QSLs
\cite{kappathermo1,131dmit2,131dmitthermo1,131dmitthermo2}.  In
particular, the spin susceptibility $\chi$ and linear coefficient
of specific heat $\gamma = c_v/T$ in $\mathrm{Ba_3NiSb_2O_9}$ both
saturate to constants at low temperature \cite{banisbo}.

Most theoretical approaches to QSLs rely on slave particle
methods, and/or wave functions which correspond to slave
particles.  While these approaches have been extensively developed
for $S=$1/2 systems, there has been little theoretical work on
them for the $S=$1 case.  We consider this here.  To sharpen the
discussion, we assume the presence of SU(2) spin symmetry, and
seek QSL states in this framework which match the basic
phenomenology so far observed in the low temperature
thermodynamics.

One way of studying spin-1 system is by introducing three flavors
of fermionic spinon $f_\alpha$ ($\alpha = 1 - 3$) as follows
\cite{ng2010,serbyn}: $\hat{S}^a = f^\dagger_{\alpha}
S^a_{\alpha\beta} f_\beta$, and $S^a$ are three spin-1 matrices.
In order to guarantee the equivalence of the spin Hilbert space
and the spinon Hilbert space, one must impose the gauge constraint
$\sum_\alpha f^\dagger_{i,\alpha}f_{i,\alpha} = 1$, fixing the
spinon density locally to 1/3-filling. At the mean field level,
the spinon $f_\alpha$ forms a Fermi surface whose area is 1/3 of
the Brillouin zone. A spinon Fermi surface seems to be consistent
with constant $\chi$ and $\gamma$ observed experimentally.
However, beyond the mean field theory, due to the single occupancy
constraint, the spinon fermi surface is coupled to a dynamical
U(1) gauge field. This U(1) gauge field has a ``dressed"
over-damped $z = 3$ dynamics due to its coupling to the Fermi
surface, which leads to a $\gamma = C_v/T \sim T^{-1/3}$ at low
temperature \cite{spinliquid1,spinliquid2}, inconsistent with
experiment.  One solution of this problem is to introduce pairing
of the spinons in the mean field state.  This has its own
difficulties: either a gap is induced and impurities must be
invoked to restore the proper
thermodynamics,\cite{groverspinliquid} or spin-rotational symmetry
must be strongly broken.\cite{serbyn}

{\it General Formalism}

We start instead by representing the spin-1 operators in the following
way:
\begin{eqnarray} \hat{S}^\mu_i = \frac{1}{2} \sum_{\alpha, \beta = \uparrow,
  \downarrow} \sum_{a = 1, 2} f^\dagger_{\alpha, a, i}
\sigma^\mu_{\alpha\beta} f^{\vphantom\dagger}_{\beta, a, i}.
\end{eqnarray}
Here $\sigma^\mu$ are three spin-1/2 Pauli matrices. Each spinon
$f_{\alpha,a}$ has two indices: $\alpha = \uparrow, \ \downarrow$
denotes spin and $a = 1, 2$ is an ``orbital'' quantum number.  Thus we
can consider not only the usual spin SU(2) rotations in the
$\alpha-\beta$ space, but also orbital SU(2) transformations in the
$a-b$ space.  Matching with the spin Hilbert space requires not only
constraining the total fermion number to {\it half-filling} (two
fermions per site), but also requiring each site to be an orbital SU(2)
singlet, which guarantees that the spin space is a symmetric spin-1
representation:
\begin{eqnarray} && \hat{n}_i = \sum_{a = 1, 2}
  \sum_{\alpha = \uparrow, \downarrow} f^\dagger_{\alpha, a, i}
  f_{\alpha, a, i} = 2,\nonumber \\ && \hat{\tau}^\mu = \sum_{\alpha, a,
    b} f^\dagger_{\alpha, a, i} \tau^\mu_{ab} f_{\alpha, b, i} =
  0. \label{constraint}
\end{eqnarray}
Here $\tau^\mu_{ab}$ are three Pauli matrices that operate on the
orbital indices. A similar slave fermion formalism with orbital
indices was introduced in Ref.~\cite{sachdev1989}, and it was
applied to two-orbital SU($N$) magnets that can be realized in
Alkaline earth cold atoms \cite{xusun,xusun2,hermelesun2}.

Due to these two independent constraints in Eq.~\ref{constraint},
the spinon $f_{\alpha,a}$ appears to have the following
$\mathrm{U(1) \times SU(2)} $ gauge symmetries:
\begin{eqnarray}
\mathrm{U(1)}_c &:& f_{\alpha, a, i} \rightarrow e^{i\theta_i}
f_{\alpha, a, i}; \cr\cr \mathrm{SU(2)}_o &:& f_{\alpha, a, i}
\rightarrow [e^{i\vec{\theta}_i \cdot \vec{\tau}/2} ]_{ab}
f_{\alpha, b, i}.
\end{eqnarray}
By rewriting $f_{\alpha,a,i}$ in terms of Majorana fermions $\eta$ as
follows, however, a larger gauge symmetry is exposed:
\begin{eqnarray}
  f_{\alpha,a,i} = \frac{1}{2}(\eta_{\alpha,a,1,i} +
  i\eta_{\alpha,a,2,i}).
\end{eqnarray}
On every site, $\eta_i$ has in total three two-component spaces, making
the maximal possible transformation on $\eta_i$ SO(8).  Within this
SO(8), the spin SU(2) transformations are generated by the three
operators $(\sigma^x\lambda^y, \ \ \sigma^y, \ \ \sigma^z \lambda^y )$,
where the Pauli matrices $\lambda^a$ operate on the two-component space
$(\mathrm{Re}[f], \mathrm{Im}[f])$. The total gauge symmetry on $\eta$
is the maximal subgroup of SO(8) that commutes with the spin-SU(2)
operators.  This is $\mathrm{Sp(4)\sim SO(5)}$ generated by the ten
matrices $\Gamma_{ab}=\frac{1}{2i} [\Gamma_a, \Gamma_b]$, where
\begin{eqnarray}
&& \Gamma_{1} = \sigma^y \tau^y \lambda^z, \ \ \Gamma_{2} =
\sigma^y \tau^y \lambda^x, \ \ \Gamma_{3} = \tau^y \lambda^y, \cr
\cr && \Gamma_4 = \tau^z, \ \ \Gamma_5 = \tau^x.
\end{eqnarray}
These $\Gamma^a$ with $a = 1 \cdots 5$ define five gamma matrices
that satisfy the Clifford algebra $\{\Gamma_{a}, \Gamma_b\} =
2\delta_{ab}$. $\Gamma_{ab}$ and $\Gamma_a$ are all $8\times 8$
hermitian matrices. $\Gamma_{ab}$ are all antisymmetric and
imaginary, while $\Gamma_a$ are symmetric.

We consider a spin-1 Heisenberg model on the triangular lattice
with both nearest neighbor and 2nd neighbor antiferromagnetic
couplings. Based on the above spinon representation of spin-1
operators, the Heisenberg model can be rewritten as follows:
\begin{eqnarray}
\sum_{i, j, \mu} J_{ij} \hat{S}^\mu_{i}\hat{S}^\mu_j &\sim&
\sum_{i, j, \mu} J_{ij} f^\dagger_{\alpha, a, i}
\sigma^\mu_{\alpha\beta} f_{\beta, a, i} f^\dagger_{\gamma, b, j}
\sigma^\mu_{\gamma \rho} f_{\rho, b, j} \cr\cr &\sim& - 2J_{ij}
\hat{\Delta}^\ast_{ab, ji} \hat{\Delta}_{ba, ji}   +
\mathrm{Const} , \cr\cr \hat{\Delta}_{ab, ji} &=&
\varepsilon_{\alpha\beta}f_{\alpha, a, j}f_{\beta, b, i}.
\label{spinrep}
\end{eqnarray}
Decoupling through a hopping term is also possible, but we do not pursue
this here. To analyze Eq.~(\ref{spinrep}), we adopt a mean field ansatz
with nonzero pairing $\langle \hat{\Delta}_{ab, ji} \rangle$, so that
the spinon $f_{\alpha,a}$ fills a mean field band structure.  To improve
beyond mean field, a variational spin wave function may be obtained by
projecting the mean field ground state to satisfy
Eq.~(\ref{constraint}):
\begin{eqnarray} | G_{\mathrm{spin}} \rangle =
\prod_i \mathrm{P}(\hat{n}_i = 2)\otimes
\mathrm{P}(\hat{\tau}^\mu_i = 0) | f_{\alpha,a} \rangle.
\label{projection}
\end{eqnarray}

The general formalism discussed above can describe many novel spin
liquid states, with various different gauge fluctuations that are
subgroups of Sp(4).  Here we focus on simple states which satisfy
the phenomenology of $\mathrm{Ba_3NiSb_2O_9}$\cite{banisbo}, and
in particular demand linear specific heat and constant
susceptibility.  We consider the following ansatz, which is a
$d+id$ pairing state of spinons:
\begin{eqnarray} \langle \hat{\Delta}_{ab,
(i,i+\hat{e}) } \rangle = \left( \delta_{ab}\Delta^{(m)}_1 +
\tau^z_{ab}\Delta^{(m)}_2\right) (e_x+ie_y)^2 , \label{ansatz}
\end{eqnarray}
where $\hat{e}$ is any of the nearest-neighbor or 2nd neighbor unit
vectors, and $\Delta^{(m)}$ with $m = 1, 2$ denotes the pairing
amplitude on the nearest and 2nd neighbor links respectively. This is a
spin singlet but orbital triplet. Because the pair wave function
vanishes when two spinons are on the same site, such states may be
particularly insensitive to the projection in Eq.~\ref{projection}.

{\em Continuum theory:}
In the majority of the paper, we consider the case with $\Delta^{(m)}_2
= 0$. Then, expanded at $\vec{k} = 0$, the low energy mean field
Hamiltonian reads
\begin{eqnarray} H &\sim& \eta^t \{ (\partial_x^2 -
\partial_y^2) \Gamma_{13} +
2 \partial_x \partial_y \Gamma_{23} \} \eta, \cr \cr  &&
\Gamma_{13} = - \sigma^y\lambda^x, \ \ \Gamma_{23} =
\sigma^y\lambda^z. \label{MF}
\end{eqnarray}
This mean field Hamiltonian has quadratic band-touching at $\vec{k} =
0$. Using the same method as introduced in Ref.~\cite{wenpsg}, one can
verify that this mean field Hamiltonian breaks the Sp(4) gauge symmetry
down to a $\mathrm{U(1)}\times Z_2$ gauge symmetry:
\begin{eqnarray} &&
  \eta_i \rightarrow e^{i\theta_i \Gamma_{45}}\eta_i, \cr \cr && \eta_i
  \rightarrow Q_i \eta_i, \ \ \ Q_i \in \{\mathbf{1}, \ \Gamma_4\}.
\end{eqnarray}
Notice that the U(1) and $Z_2$ gauge transformations do {\it not}
commute with each other.

In addition to the quadratic band touching at $\vec{k} = 0$, depending
on $\Delta^{(m)}$, there are multiple Dirac points in the Brillouin
zone. For instance, when $\Delta^{(2)} < \Delta^{(1)}$, there are Dirac
points at the Brillouin zone corners $\vec{Q} = \pm (4\pi/3, 0)$. A
complex Dirac fermion field $\psi$ at momentum $\vec{Q} = (4\pi/3, 0)$
can be defined as
\begin{eqnarray} \eta_{\vec{r}} = \psi_{ \vec{r}}
e^{i\vec{Q}\cdot \vec{r}} + \psi^\dagger_{\vec{r}} e^{-i\vec{Q}\cdot
  \vec{r}}.
\end{eqnarray}
The low energy Hamiltonian for $\psi$ reads $H_\psi \sim \psi^\dagger (i
\Gamma_{13}\partial_x - i\Gamma_{23}\partial_y) \psi $.  However, the
Dirac fermion has vanishing density of states at zero energy, and thus
contributes sub-dominantly to the $\eta$ spinon in many physical
properties.

The spinon carries a projective representation of physical symmetry
group. Under discrete symmetry transformations, the low energy spinon
fields $\eta$ and $\psi$ transform as:
\begin{eqnarray}
  T_x &:& x \rightarrow x+1,
  \ \ \ \ \eta \rightarrow \eta, \ \ \ \ \psi \rightarrow e^{i4\pi/3}
  \psi; \cr\cr T &:& t \rightarrow -t, \ \ \ \ \eta \rightarrow i
  \Gamma_{12} \eta, \ \ \ \ \psi \rightarrow i\Gamma_{12} \psi^\dagger;
  \cr\cr \mathcal{I} &:& \vec{r} \rightarrow - \vec{r}, \ \ \ \ \eta
  \rightarrow \eta, \ \ \ \ \psi \rightarrow \psi^\dagger; \cr\cr
  \mathrm{P}_y &:& x \rightarrow -x, \ \ \ \ \eta \rightarrow i\Gamma_{13}
  \eta, \ \ \ \ \psi \rightarrow i\Gamma_{13} \psi^\dagger; \cr\cr R_{\pi/3}
  &:& (x+iy) \rightarrow e^{i\pi/3}(x+iy), \cr \cr && \eta \rightarrow
  e^{i\frac{\pi}{3} \Gamma_{12}} \eta, \ \ \ \ \psi \rightarrow
  e^{i\frac{\pi}{3} \Gamma_{12}}\psi^\dagger .
\end{eqnarray}
These transformations guarantee that there is no relevant fermion
bilinear perturbation that does not break physical symmetry. For
instance, the fermion bilinear $f^\dagger_{i}f_i \sim \eta^t \Gamma_{12}
\eta$ breaks the time-reversal symmetry; thus it is forbidden in the
Hamiltonian.

{\em Effect of gauge fluctuations:}
The spinons are coupled to a U(1) gauge field $a_\mu \Gamma_{45}$.  The
gauge field Lagrangian is renormalized by the fermion loop, which
generates a mass gap for $a_0$, thus $a_0 $ can be ignored
hereafter. The same fermion loop will also renormalize the dynamics of
the transverse component of gauge field $a^T$ to be:
\begin{eqnarray}
  \mathcal{L}_1 &=& \sum_{\omega, \vec{q}} \left( c |\omega| +
    \frac{q^2}{e_{\omega, \vec{q}}^2} + c_1\sqrt{\omega^2 + v^2q^2}
  \right)|a^T_{\omega,\vec{q}}|^2, \cr \cr e^2_{\omega, \vec{q}} &\sim&
  \frac{e^2}{1 + c_2 e^2 \log\left(\frac{\Lambda^2}{4\omega^2 +
        q^4}\right)}. \label{qqed}
\end{eqnarray}
In Lagrangian $\mathcal{L}_1$, the first two terms come from the
screening of spinons at the quadratic band touching, while the
third term comes from the Dirac points.   At low energy, the gauge
field therefore obeys $z=1$ scaling with $\omega \sim q$, so that
the $q^2/e^2$ term is negligible in Eq.~(\ref{qqed}).  For this
reason, the gauge field decouples from $\eta$ (which has $z=2$
scaling) at low energy, but remains strongly coupled to the $z=1$
Dirac fermion $\psi$.

{\em Thermodynamic and transport properties:}
The finite density of states of the $\eta$ spinon leads to a constant
$\gamma = C_v /T$ at zero temperature. In terms of $\eta$, the spin
density $S^z$ is represented as
\begin{eqnarray} \sum_i S^z_i = \sum_{\alpha, a, i}
\eta^\dagger_{i} \sigma^z \lambda^y \eta_{i}.
\end{eqnarray}
Since the spin density commutes with the mean field Hamiltonian
Eq.~\ref{MF}, turning on an external magnetic field creates a Fermi
surface of $\eta$ and $\psi$, and since the density of states is finite
at the quadratic band touching, the spin susceptibility saturates to a
constant at zero temperature. Thus this spin liquid phase is consistent
with the scaling of specific heat and spin susceptibility observed
experimentally.

We also find a unique scaling of thermal conductivity. The thermal
conductivity is proportional to the specific heat and the velocity of
entropy carriers of the system: $\kappa \sim c v l $. Since the spinon
has quadratic band touching, at low temperature the average velocity
scales as $v \sim T^{1/2}$, thus the thermal conductivity scales as
$\kappa \sim T^{3/2}$.  Note that the thermal conductivity contribution
from the Dirac fermion and gauge field scales  as $T^2$,
and is thus subdominant.

{\em Fluctuating orders:} The gauge invariant fermion bilinear operators
can be viewed as physical order parameters with power-law
correlations. They can be classified according to their transformations
under symmetry and gauge symmetry. Some of the fermion bilinears are
summarized as follows:
\begin{eqnarray} (1) && \mathrm{Spin \ density \ wave} \cr\cr &&
  \vec{S}_{\vec{r}} = \cos(\vec{Q}\cdot \vec{r})\vec{n}_1 +
  \sin(\vec{Q}\cdot \vec{r})\vec{n}_2, \cr\cr && \vec{n}_1 + i\vec{n}_2
  \sim a_1 \psi^\dagger \vec{S} \psi^* + b_1 \eta^t \vec{S}
  \psi, \cr\cr  (2) && \mathrm{Spin-Nematic} : \cr\cr && \vec{d} \sim a_2
  \eta^t \vec{S} \Gamma_3 \eta + b_2 \psi^\dagger \vec{S} \Gamma_3 \psi,
  \cr\cr (3) && \mathrm{Nematic}: \cr\cr && N = \sum_{\hat{e}} \vec{S}_i
  \cdot \vec{S}_{i+\hat{e}} (e_x + i e_y)^2 \cr\cr && N = N_1 + i N_2
  \cr\cr &\sim& a_3(\eta^t \Gamma_{13} \eta + i \eta^t \Gamma_{23} \eta)
  + b_3 (\psi^\dagger \Gamma_{13} \psi + i \psi^\dagger \Gamma_{23}
  \psi); \cr\cr (4) &&
  \mathrm{Spin-chirality} : \cr\cr && C = \sum_{ijk \in \bigtriangleup}
  \vec{S}_i \cdot(\vec{S}_j \times \vec{S}_k) + \cdots \cr\cr && C \sim
  a_4 \eta^t \Gamma_{12} \eta + b_4 \psi^\dagger \Gamma_{12} \psi,
  \label{order} \end{eqnarray}
where $\vec{S}=(\sigma^x\lambda^y, \sigma^y,\sigma^z \lambda^y)$
are the spin matrices.

\begin{figure}
\begin{center}
\includegraphics[width=3.2 in]{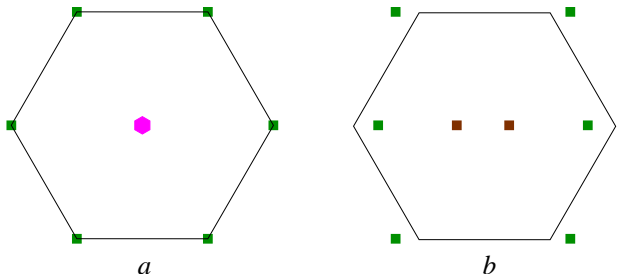}
\caption{$a$, The spin liquid we are considering contains a
quadratic band touching at $\vec{k} = 0$ (hexagon), and Dirac
points (squares) at the corners of the Brillouin zone. $b$, with a
nonzero and small nematic order $N_1 > 0$, the quadratic band
touching is split into two Dirac points, and the locations of the
other Dirac points are shifted.} \label{trianglespin1}
\end{center}
\end{figure}

Several of these orders are germane to spin-one triangular
antiferromagnets.  The spin density wave order parameter is
precisely that which describes the classical 120$^\circ$ planar
spin state, with $\vec{Q}=(4\pi/3,0)$ coinciding with the
Brillouin zone corner.  As a consequence the spin structure factor
of this state is singular at this momentum. Spin nematic order
occurs naturally when biquadratic interactions are present in spin
one systems.\cite{nitheory1} In fact, $\vec{d}$ changes sign under
the $Z_2$ gauge transformation $\eta \rightarrow \Gamma_4 \eta$,
so it is a headless nematic director. The physical order parameter
is actually a bilinear of $\vec{d}$, which corresponds to the
ferro-quadrupoletensor
\begin{eqnarray} Q^{\mu\nu} = \frac{1}{2} \langle \hat{S}^\mu_i
  \hat{S}^\nu_i + \hat{S}^\nu_i \hat{S}^\mu_i\rangle -
  \frac{2}{3}\delta^{\mu\nu} = d^\mu d^\nu -
  \frac{|\vec{d}|^2}{3}\delta^{\mu\nu}.
\end{eqnarray}
Spatial nematic order, in which lattice rotation symmetry is broken but
time-reversal and spin symmetry are preserved, is described by $N_1$ and
$N_2$.  Order of this type was suggested for S=1 triangular
antiferromagnets in Ref.\cite{balentstrebst}, but also can be realized
by spontaneous formation of Haldane chains.  The spin-chirality order
parameter $C$ is less obvious from a microscopic perspective, but is a
fluctuating order for this QSL state.

At the mean field level, the equal time correlation of spin
chirality, nematic, and spin density wave order parameters all
fall off as $1/r^4$; the correlation of spin quadrupole order
parameter falls off as $1/r^8$. The U(1) gauge fluctuation will
modify the scaling dimension of the order parameters, and its
correction can be calculated systematically using a $1/N$
expansion. We will leave this calculation to future studies.

{\em Potential instabilities:} One potential instability of this
spin liquid state is instanton (monopole) proliferation of the
compact U(1) gauge field. However, due to screening by the gapless
fermions, the instantons are greatly suppressed.  By analogy with
the theory of the algebraic spin liquid \cite{hermelemonopole} (in
which the $z=1$ gauge field is similarly strongly coupled to Dirac
fermions), we expect the spin liquid phase here to be similarly
stable in principle.

Furthermore, the mean field Hamiltonian Eq.~\ref{MF} is subject to
perturbations such as four-fermion interactions, which are
marginal perturbations at the quadratic band touching according to
naive power-counting. These four-fermion interactions can modify
the correlation functions of the order parameters discussed above.
The renormalization group may lead to weak run-away flow of these
four-fermion interactions, which eventually can break the symmetry
of the system, and develop one of the orders in Eq.~\ref{order}.

If one of these orders develops, it can completely or partially gap the
fermions and introduce interesting effects.  Nonzero spin nematic order,
$\vec{d}\neq 0$, gaps out the quadratic band touching and Dirac fermion
$\psi$.  Depending on the sign of $a_2$ and $b_2$, a nonzero $\vec{d}$
drives the mean field band structure of spinon into either a quantum
spin Hall type of topological insulator or a topologically trivial
insulator. If the system is in a quantum spin Hall topological
insulator, assuming $\vec{d}$ is ordered along $\hat{z}$ direction, the
quantized flux of U(1) gauge field $a_\mu \Gamma_{45}$ would carry spin
$S^z$. Since the 2+1d photon phase of the U(1) gauge field is the
condensate of gauge flux, the U(1) spin rotation around $\hat{z}$ axis
is spontaneously broken in the photon phase, thus eventually the spin
SU(2) symmetry is broken down to a discrete subgroup $i.e.$ there are in
total three Goldstone modes instead of two. If the spinon band insulator
has trivial topology, then the system is in an ordinary
ferro-quadrupolar phase as discussed in
Ref.~\cite{senthilni,balentstrebst}.

Weak spatial nematic order does not open a gap but only splits the
quadratic band touching into Dirac fermions at two different momenta
(Fig.~\ref{trianglespin1}); the original Dirac fermions $\psi$ also
shift. When the nematic order magnitude is very strong, above some
critical value, all the Dirac fermions meet and annihilate in pairs, and
the spinons become fully gapped.

Spin chirality order, which breaks time-reversal and reflection
symmetries, gaps out both the quadratic band touching and the
Dirac points. Depending on the sign of $a_4$ and $b_4$, a nonzero
spin chirality order can drive the spinons into a topological
Chern insulator, or a topologically trivial band insulator with
the same symmetry. In the former case, one obtains a chiral spin
liquid, in which the U(1) gauge field $a_\mu \Gamma_{45}$ acquires
a Chern-Simons term after integrating out the fermions.  In the
topologically trivial band insulator, the U(1) gauge field will
become confined by instanton proliferation.

{\em Other phases:} For $S=1$ spins, we may also consider another state
with $\Delta^{(m)}_1 $ and $\Delta^{(m)}_2$ both nonzero, and
$|\Delta^{(m)}_1| \neq |\Delta^{(m)}_2|$. In this case, the spinons have
two different bands both with quadratic band touching at $\vec{k} = 0$,
but they have different band curvature:
\begin{eqnarray} H &\sim& \eta^t
  \{ (\partial_x^2 - \partial_y^2) (A \Gamma_{13} + B \Gamma_{25})
  \cr\cr &+& 2 \partial_x \partial_y (A \Gamma_{23} - B \Gamma_{15}) \}
  \eta. \label{MF}
\end{eqnarray}
$A$ and $B$ are two linear combinations of pairing amplitudes on nearest
and 2nd neighbor links. In this state, the gauge symmetry is broken down
to $Z_4$:
\begin{eqnarray} \eta_i \rightarrow Q_i \eta_i, \ \ \ Q_i \in
  \{\pm 1, \ \pm \Gamma_4\}.
\end{eqnarray}
The $Z_4$ gauge field has a deconfined phase in 2+1 dimension, and this
state is thus clearly locally stable. It also exhibits non-zero finite
spin susceptibility and $\gamma = C_v/T$ at zero temperature.

A similar $d + id$ state with quadratic band touching can also be
considered for spin-1/2 systems on the triangular lattice. The
same mean field Hamiltonian as Eq.~\ref{MF} applied, but without
an orbital index. This state remains time-reversal and reflection
invariant, and has $Z_2$ gauge structure.  One might consider this
as a candidate state for the spin liquids observed in the
compounds $\kappa\mathrm{-(ET)_2Cu_2(CN)_3}$,
$\mathrm{EtMe_3Sb[Pd(dmit)_2]_2}$, and $\mathrm{Ba_3CuSb_2O_9}$
\cite{kappa1,kappa2,131dmit1,131dmit2,131dmit3,bacusbo}.

{\em Future work:} In Ref.~\cite{ran2007,groverspinliquid}, a
variational Monte Carlo computation based on the Gutzwiller
projected wavefunction for $S=1/2$ was used to compare the energy
of various mean field spin liquid states. Extension of this method
to our generalized $S=1$ projected wavefunction,
Eq.~\ref{projection}, is a non-trivial and interesting problem for
the future.

We would like to acknowledge support from the NSF under grants
DMR-1101912 (Fisher) and DMR-0804564 (Balents). Xu is supported by
the Sloan Research Fellowship.

\bibliography{triangle}

\end{document}